\documentclass[acmsmall]{acmart}

\pdfoutput=1

\usepackage{multirow}
\usepackage{graphicx}
\usepackage{caption}
\usepackage{float}
\usepackage{subcaption}
\usepackage{array}
\usepackage{makecell}
\usepackage{xcolor}
\usepackage{soul}
\usepackage[normalem]{ulem}
\usepackage[most]{tcolorbox}

\AtBeginDocument{%
  }

\acmISBN{978-1-4503-XXXX-X/2018/06}

\begin{document}

\title{From OSS to Open Source AI: an Exploratory Study of Collaborative Development Paradigm Divergence}


\author{Hengzhi Ye}
\email{hzye@stu.pku.edu.cn}
\orcid{0009-0003-2892-5726}
\affiliation{%
  \institution{School of Computer Science, Peking University}
  \city{Beijing}
  \state{Beijing}
  \country{China}
}

\author{Minghui Zhou}
\email{zhmh@pku.edu.cn}
\orcid{0000-0001-6324-3964}
\authornote{Corresponding author.}
\affiliation{%
  \institution{School of Computer Science, Peking University}
  \city{Beijing}
  \state{Beijing}
  \country{China}
}

\renewcommand{\shortauthors}{Hengzhi Ye and Minghui Zhou}

\begin{abstract}
AI development is embracing open-source paradigm, but the fundamental distinction between AI models and traditional software artifacts may lead to a divergent open-source development paradigm with different collaborative practices, which remains unexplored.
We therefore bridge the knowledge gap by quantifying and characterizing the differences in the collaborative development paradigms of traditional open source software (OSS) and open source AI models (OSM), and investigating the underlying factors that may drive these distinctions.
We collect 1,428,792 OSS repositories from GitHub and 1,440,527 OSM repositories from HF Hub, and conduct comprehensive statistical, social network and content analyses to measure and understand the differences in collaboration intensity, collaboration openness, and user innovation across the two development paradigms, complementing these quantitative results with semi-structured interviews.
In consequence, we find that compared to OSS development paradigm, the OSM development paradigm exhibits significantly lower collaboration intensity; lower collaboration openness regarding direct contribution while persisting relatively open knowledge exchange; and a divergence toward adaptive utilization user-innovation rather than collaborative improvement. Through semi-structured interviews, we further elucidate the socio-technical factors underlying these differences.
These findings reveal the paradigmatic divergence in open source development between traditional OSS and OSM across three critical dimensions of open source collaboration and potential underlying factors, shedding light on how to improve collaborative work techniques and practices within the context of AI development.
\end{abstract}

\setcopyright{cc}
\setcctype{by}
\acmJournal{PACMHCI}
\acmYear{2026} \acmVolume{10} \acmNumber{6} \acmArticle{CSCW090}
\acmMonth{10} \acmDOI{10.1145/3816938}



\received{May 13, 2025}
\received[revised]{January 13, 2026}
\received[accepted]{April 9, 2026}

\maketitle

\section{Introduction}\label{sec:intro}

Open Source Software (OSS) development has emerged not merely as a technical methodology, but as a fundamental paradigm of collaborative intelligence~\cite{leimeister2010collective} characterized by collaborative participation, inclusive processes, and user innovation~\cite{hippel2003open, chikersal2017deep}. This paradigm, conceptualized through Raymond's \textit{``bazaar''} metaphor~\cite{raymond1999cathedral}, attracts developers with genuine interest and passion for specific projects, transcending geographical and temporal barriers to facilitate flexible collaboration among contributors distributed worldwide, ultimately yielding products of superior functionality and quality~\cite{o1999lessons, mockus2002two}. However, the open source ethos underlying this success may raise critical questions about its adaptation and manifested development paradigm when the collaborative artifact changes.

In recent years, Artificial Intelligence has experienced unprecedented advancement, drawing substantial attention and investment from researchers and commercial enterprises. The development of AI models is increasingly embracing open-source paradigm: companies engaged in AI development increasingly demonstrate a preference for open-sourcing their models~\cite{vake2025open, latentSpace}, and concurrently, numerous smaller collectives and individual developers have shared their models with the global community, establishing this as a defining trend of the current technological landscape~\cite{osborne2024ai, jiang2023empirical}. Here, we use the term Open Source AI Model (OSM) to refer to AI model whose components, including but not limited to architecture, trained weights, and associated documentation, are made publicly available under licenses that permit use, modification, and redistribution~\cite{white2024model}\footnote{Note that though ideal open source ethos advocates complete transparency of all components in collaborative artifacts, many AI models' underlying training data and complete training process may not always be fully accessible or reproducible. Despite limiting free reproduction and distribution, we consider such models as part of OSM in this study, given the prevalence of this phenomenon in current AI development.}.
In fact, the contemporary AI ecosystem is flourishing increasingly due to the benefits derived from OSM development paradigms. Practitioners fine-tune base models to produce domain-specific ones suitable for downstream tasks~\cite{sun2019fine, wortsman2022robust}, adapt structural components from exemplary models~\cite{zhao2020combining}, and utilize publicly available datasets for model training~\cite{bhaskaran2023enhanced}. These development practices, facilitated by open source, have contributed significantly to the vibrant AI development ecosystem.

The emergence of OSM presents an adaption of the open-source paradigm to fundamentally different artifacts{~\cite{aiDifferent}}, prompting discussions within both academic and practitioner communities about whether traditional OSS collaborative practices apply to AI model development. Existing academic studies and online community discussions have explored the conceptual aspect of differences between OSS and OSM (i.e., the collaborative artifact itself), suggesting that the concept of ``open source'' applies differently to each domain~\cite{widder2024open, aiDiffoss, aiDifferent, vake2025open}. Accordingly, the distinction primarily lies in the divergence from the sharing of building logic within source code in OSS to the disclosure of training results in OSM. In practice, this often manifests as OSM primarily sharing model architectures and weights more than just executable code, yet appearing less transparent than OSS due to probable omission of training data and process recipes.

While for the CSCW community, since the divergence of collaborative artifact behind the ``open source'' may fundamentally reshape the socio-technical orientation of development, a more pressing concern involves understanding whether it leads to different open source development paradigms, especially how collaborative practices occur through online platform, because such variations impact developer experiences, development efficiency, and product outcomes. It is intuitive to recognize that open source paradigms applied to different subjects often yield distinct manifestations and behaviors. However, despite such widespread recognition that ``something different is happening'', no systematic empirical study has quantified these differences, established their scope and magnitude, or elucidated the underlying factors.

To address the knowledge gap, we try to answer the core question: do the OSS and OSM development paradigms exhibit measurably different collaborative behaviors? Specifically, we characterize the collaborative paradigm divergence between OSS and OSM development and investigate the underlying factors driving these differences, focusing on collaboration intensity, collaboration openness and innovation patterns reflected by community communication messages, which represent critical dimensions of open source development paradigm (i.e., collaborative participation, inclusive processes and user innovation)~\cite{hippel2003open}, leading to the following research questions (RQs):

\begin{itemize}

\item \textbf{RQ1}: How does collaboration intensity differ between OSS and OSM development paradigms?

\item \textbf{RQ2}: How does collaboration openness differ between OSS and OSM development paradigms?

\item \textbf{RQ3}: How does user innovation differ between OSS and OSM development paradigms? 

\item \textbf{RQ4}: What are the potential socio-technical factors underlying the differences between OSS and OSM development paradigms?

\end{itemize}

Following these research questions, we conducted a mixed-methods study combining large-scale quantitative analysis with qualitative semi-structured interviews. Specifically, we extensively collected and analyzed activity data from 1,428,792 OSS repositories on GitHub and 1,440,527 OSM repositories on Hugging Face (HF) Hub. We found three key differences. First, intensity of collaboration in OSM is significantly lower than in OSS, with interaction metrics in OSS exceeding OSM by over 100-fold and collaboration network showing higher connectivity in OSS communities. Second, OSM collaboration demonstrates limited openness, with 98.91\% of contributors affiliated with publishing organizations or HF staff and high overlap (49.7\%) between collaborative clusters and formal organizations. Third, communication in OSS community focuses on collaborative improvement (e.g., \textit{bug reports}, \textit{feature suggestions}), whereas that in OSM community emphasizes adaptive utilization (e.g., \textit{usage problems}, \textit{performance evaluation}), which reflects a change in user innovation pattern from directly contributing to artifacts in OSS to applying and adapting models in OSM. Complementing these findings, semi-structured interviews further elucidate potential underlying factors driving these distinctions, including architectural barriers, resource constraints, infrastructural misalignment and corporate strategic factors.

In Section~\ref{sec:relatedwork} we review related work. We conduct a preliminary analysis of OSS and OSM development pipelines to prompt subsequent quantitative analysis in Section~\ref{sec:pre}. Section \ref{sec:method} presents the research methodology and Section \ref{sec:result} reports the results. We discuss the implications and limitations in Section \ref{sec:discuss} and conclude in Section \ref{sec:conclude}.

\section{Related Work}\label{sec:relatedwork}

\subsection{Lessons of OSS development}\label{subsec:OSSdev}

OSS has shown consistent growth and increasing popularity within the software ecosystem since the free software movement of the last century~\cite{stallma2006free}. Raymond~\cite{raymond1999cathedral} metaphorically characterized this open-source development approach as a \textit{``bazaar''}, contrasting it with the \textit{``cathedral''} model of traditional commercial development to emphasize the collaborative advantages of open-source paradigm. As OSS entered the mainstream market at the beginning of this century, it garnered substantial academic attention. Researchers began systematically examining the ascendance trajectory of OSS, identifying factors contributing to its success, and developing frameworks to understand its ecosystem~\cite{wu2001open, almarzouq2005open}. These investigations revealed that the flexibility and agility of OSS and effective utilization of collective intelligence~\cite{chikersal2017deep, leimeister2010collective} contribute to the advantage of OSS.

Considering the development process, Mockus et al.~\cite{mockus2002two} focused their research on the development and maintenance processes of two prominent OSS projects, illuminating common patterns in personnel structure, architectural organization, contributor distribution, and operational mechanisms within successful OSS projects. Paulson et al.~\cite{paulson2004empirical} conducted an empirical comparative study between open-source and closed-source software development and found that creativity is more prevalent in OSS development processes. Subsequent researches have continued to examine the evolving landscape of OSS development, providing valuable insights into how OSS development adapts to changing technological landscapes, integrates into new ecosystems, and continues to refine its practices~\cite{fitzgerald2006transformation, zhou2012make}. 

These researches together revealed that the OSS paradigm's success derives from its three foundational dimensions working synergistically. First, distributed participation enables distributed contributors to coordinate their efforts and collective intelligence~\cite{hippel2003open, dabbish2012social}. Second, inclusive processes ensure that both source code and development processes remain accessible for inspection, modification, and redistribution, facilitating knowledge transfer and collective problem-solving~\cite{almarzouq2005open}. Third, user innovation allows those who directly benefit from the software to drive its development, creating a direct feedback loop between user needs and software development~\cite{hippel2003open, raymond1999cathedral}.

The establishment of GitHub marked a pivotal moment in OSS development history. Leveraging the efficiency and superiority of the Git version control system along with the principles of social coding, which advocate transparency, openness, sharing, and collaboration, GitHub rapidly gained popularity to become developers' predominant online code hosting platform~\cite{dabbish2012social, zagalsky2015emergence, mcdonald2013performance}. Currently hosting more than 150 million developers and exceeding 420 million repositories~\cite{github}, GitHub's unparalleled popularity and massive record of various activity data have positioned it as a valuable analytical subject and data source for OSS researches~\cite{cosentino2016findings, gousios2017mining, dabic2021sampling}.

Previous research utilizing GitHub data encompasses diverse dimensions: from artifact-centric aspects such as project type and quality, compositional characteristics, version release, and documentation practices~\cite{jarczyk2014github, aggarwal2014co, wu2022demystifying}, to process-oriented elements such as platform-based communication mechanisms, issue tracking, and its resolution~\cite{ kinsman2021software, brisson2020we, bissyande2013got}. GitHub continues to facilitate nuanced understanding of contemporary OSS development practice as a rich data source.

\subsection{Emergency of OSM and its sharing}\label{subsec:OSMdev}

The open-source ethos extends well beyond traditional software development and distribution, permeating various domains of technological and knowledge sharing, including Artificial Intelligence and Machine Learning. As early as a decade ago, Sonnenburg et al.~\cite{sonnenburg2007need} advocated for OSS to support machine learning advancement, identifying the scarcity of ML-oriented OSS as a significant impediment to research advances. This early intersection between open-source ethos and AI/ML presaged a deeper convergence that would later emerge, which represents the application of open source ethos to AI models.

However, open-sourcing of AI models remained relatively uncommon for several years. This hesitancy was due to legitimate concerns regarding potential security implications of publicly releasing models and commercial considerations of AI companies weighing operational costs against potential benefits~\cite{solaiman2019release}. Things changed with transformative developments in the fields of AI in recent years, and the trend toward open-sourcing AI models has gained remarkable momentum~\cite{BenefitOSM}. Researchers and practitioners have increasingly recognized the benefits that open development methodologies bring to AI advancement~\cite{langenkamp2022open, vake2025open} as OSM has become the mainstream practice, exemplified by several landmark releases: Stability AI's \texttt{Stable Diffusion}~\cite{diffusion}, Meta's \texttt{LLaMA 2}~\cite{Llama}, and Deepseek's \texttt{deepseek-r1}~\cite{deepseek}.

Concurrent with the rise of OSM, specialized platforms dedicated to share and develop AI/ML projects have emerged, most notably HF Hub. Initially established as a hosting platform for AI/ML artifacts where people could interconnect related projects with their respective datasets, HF Hub has evolved into a GitHub-like platform specifically tailored for the AI/ML domain and gained significant popularity, with more than 1,600 thousand models and 350 thousand datasets hosted~\cite{hfhub, ait2025suitability}. HF Hub now incorporates collaborative features designed to enhance community interactions, such as discussion forums that facilitate user communication and spaces that enable immediate experimentation with model demonstrations, solidifying its position as the primary platform for sharing and developing OSM.

Several studies have analyzed how the emergence of HF Hub has accelerated OSM development and advancement in the broader AI field~\cite{jones2024we, shen2023hugginggpt}. These investigations have examined the platform's status and challenges regarding model reuse, documentation, and licensing practices~\cite{ait2023hfcommunity, castano2024lessons, jiang2023empirical}, paralleling previous researches on GitHub in the traditional OSS domain. For instance, Pepe et al.~\cite{pepe2024hugging} highlighted transparency deficiencies in model storage on HF Hub, manifested through dataset absence and licensing omissions or mismatches. Similarly, Castaño et al.~\cite{castano2024analyzing} identified problems with documentation practices, including unclear and incomplete model cards, and revealed developers' tendency to conduct model perfection rather than adaptation and correction. These findings collectively indicate immature collaborative practices within the OSM ecosystem.

\subsection{Communication and collaboration in technical development ecosystems}

Communication and collaboration form the backbone of open-source technical development, with research demonstrating their critical role in project success across various domains~\cite{tymchuk2014collaboration, yamauchi2000collaboration}. These interactive processes facilitate knowledge exchange among developers separated by geographical and temporal boundaries, enabling open-source development to effectively harness collective intelligence and build sustainable communities~\cite{raymond1999cathedral}.

In OSS ecosystems, communication often takes place through multiple channels, such as issue trackers, discussion forums, PR/commit messages, mailing lists, and repository documentation~\cite{xuan2014building, bertram2010communication, guzzi2013communication}. These diverse communication modalities collectively enable developer collaboration, knowledge sharing, and community cultivation. Research approaches to understanding these communication patterns have followed both micro and macro perspectives. From a microscopic perspective, researchers have selected representative cases to analyze communication best practices and to examine how specific interactions contribute to project outcomes~\cite{tan2019communicate, herbsleb2003empirical}. On the other hand, macroscopic analyses have investigated communication frequency and patterns across entire development ecosystems, exploring how the efficiency of communication channels, the transparency of decision-making processes, and the status of documentation influence both contribution quality and overall product excellence ~\cite{gutwin2004revealing,geiger2018types, zhang2020data}, as well as characterizing the collaboration network and community formation of developers~\cite{moradi2021community, wu2014exploring}.

GitHub's emergence as the predominant platform for OSS development and sharing has positioned communication and collaboration that occur there as exemplars of the broader OSS development process~\cite{dabbish2012social, mcdonald2013performance}. Consequently, many platform-based studies attempting to understand communication and collaboration in OSS have emerged, often focusing on PR/commit messages, issue reports, and associated discussions within the GitHub ecosystem~\cite{dabbish2012social, ortu2018mining, hata2022github}. Ortu et al.~\cite{ortu2018mining} mined data from GitHub to uncover communication patterns in OSS development processes, revealing different frequencies and content in communications between developers and users occupying different roles, while Tsay et al.~\cite{tsay2014let} evaluated contributions through discussion on GitHub, substantiating GitHub's dual functionality as both a development and social platform and highlighting the intimate relationship between communication practices and development outcomes. Furthermore, using topological analysis techniques, Tantisuwanku et al.~\cite{tantisuwankul2019topological} systematically investigated communication channels and knowledge sharing dynamics on the GitHub platform. These platform-based researches on communication and collaboration have contributed to constructing a comprehensive understanding of interactions in the OSS ecosystem.

Similarly, OSM development, as the practice of applying the open source ethos to a different artifact, inherently involves knowledge exchange and developer collaboration. Piorkowski et al.~\cite{piorkowski2021ai} investigated communication modalities and characteristics of OSM developers in a multidisciplinary team, while Christopher et al.~\cite{akiki2022bigscience} documented the globally distributed development process of \texttt{Bloom} by the BigScience workshop\cite{workshop2022bloom}. These studies primarily investigated OSM development practices by recording individual project histories and exploring communication modalities within project workflows, thereby laying the foundation for a deeper understanding of OSM collaborative patterns and development ecosystems. On the other hand, HF Hub, as the mainstream platform for hosting and sharing OSMs, captures extensive activity data from AI/ML developers, providing a rich data source~\cite{ait2025suitability} for empirical research on communication and collaboration within OSM development.

Despite significant work on open source and AI development practices, a critical knowledge gap persists regarding the collaborative development paradigm divergences between OSS and OSM development. OSS and OSM represent different collaborative artifacts under the open source ethos, and intuitive reasoning suggests that ``something different is happening''. However, systematic investigation into these differences is lacking. Specifically, what differences exist between OSM and OSS development paradigms across various dimensions, how to quantify these differences, and what underlying factors may drive such distinctions remain unexplored.

\section{Preliminary Analysis}\label{sec:pre}

In this section, we begin with a preliminary analysis of OSS and OSM development process, modeling their pipelines to identify potential differences that prompt subsequent quantitative investigation.

OSS development operates through collaborative practices that enable distributed coordination while maintaining software quality and coherence~\cite{almarzouq2005open}. These collaborative mechanisms manifest mainly across two different but interconnected phases that influence the OSS development lifecycle.

First, during the \textbf{core development} phase, internal organizational collaboration typically occurs through systematic commit submissions, enabling parallel development of discrete software components while maintaining integration integrity. This process embodies modular collaboration, which refers to the ability to decompose complex software into manageable units that can be developed independently while integrated seamlessly. Meanwhile, external developers can contribute through pull requests mechanisms, where proposed modifications undergo review before potential integration~\cite{mockus2002two, dabbish2012social}, creating a quality-gated inclusion process that balances openness with technical standards. The second phase involves the \textbf{downstream feedback}: after a software version is released, downstream users may implement incremental features based on the released version, or discover and patch vulnerabilities. These users then provide feedback to upstream developers through commits or pull requests, effectively extending the development community beyond original creators and establishing recursive collaboration patterns where today's users become tomorrow's developers~\cite{heron2013open, workinoss}. 

However, in OSM development, different artifacts may lead to fundamentally different technical constraints and development processes across both development phases. In the \textbf{core development} phase, OSM development processes critically involve data engineering beyond model architecture construction~\cite{whang2023data, sambasivan2021everyone}, which includes the complex processes of data curation, preprocessing, and quality validation. This is followed by model pre-training and subsequent fine-tuning processes on the prepared data, which require substantial computational resources and specialized expertise. Considering the \textbf{downstream engagement} phase, after an OSM is trained and released, downstream users might deploy it for inference, directly call its API, or, when domain-specific requirements arise, use the released model as a starting point for further fine-tuning until achieving desired performance~\cite{jiang2023empirical,sun2019fine,ziegler2019fine} without contributing back to the original model.

\begin{figure}[hbt]
    \centering
    \includegraphics[width=0.82\linewidth]{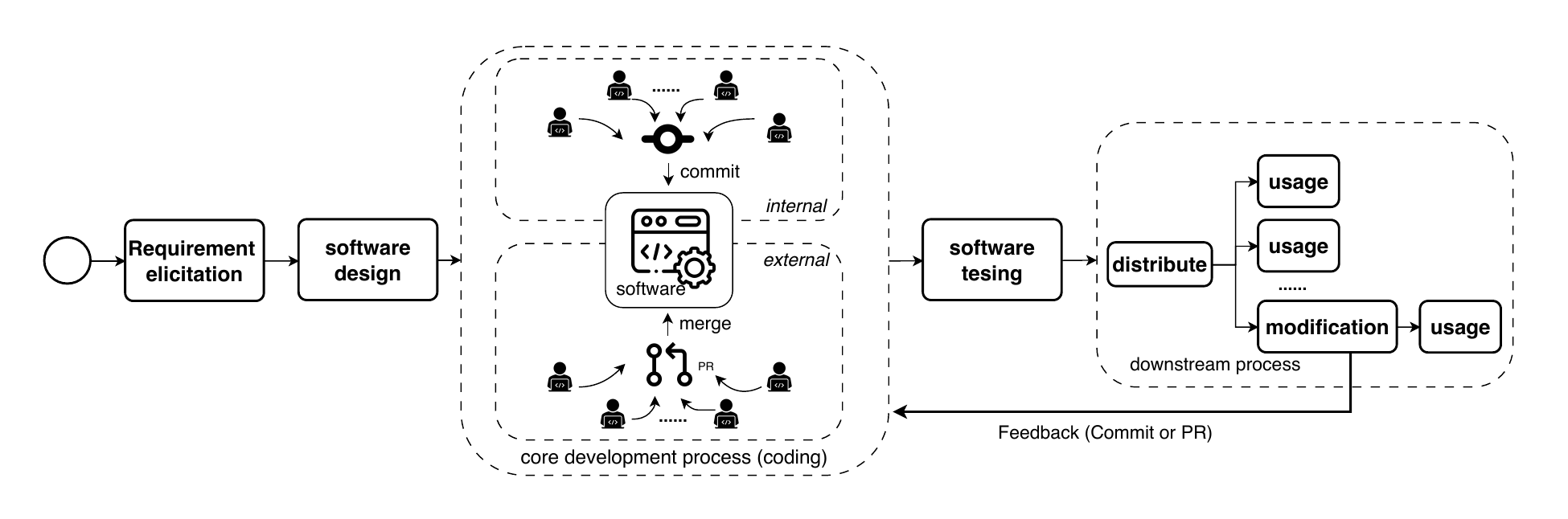}
    \vspace{-.5em}
    \caption{OSS development process pipeline}
    \label{fig:oss_pipepline}
    \Description{oss pipeline which includes requirement elicitation, software design, coding process, software testing and downstream process}
    \vspace{-1em}
\end{figure}

\begin{figure}[hbt]
    \centering
    \vspace{-.5em}
    \includegraphics[width=0.95\linewidth]{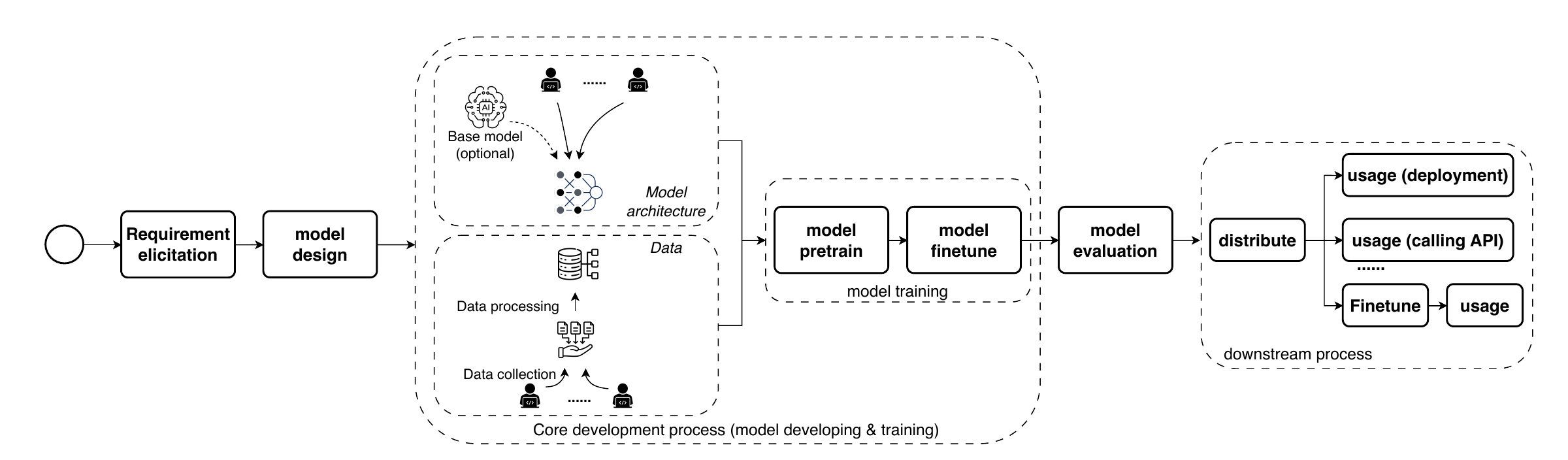}
    \vspace{-.5em}
    \caption{OSM development process pipeline}
    \label{fig:osm_pipepline}
    \Description{osm pipeline which includes requirement elicitation, model design, model developing and training phase, model evaluation and downstream process}
    \vspace{-.5em}
\end{figure}

Based on considerations above and synthesizing extensive literature~\cite{almarzouq2005open, mockus2002two, dabbish2012social, heron2013open, workinoss,whang2023data, sambasivan2021everyone,jiang2023empirical,sun2019fine}, we modeled the development processes for both open-source paradigms as pipelines illustrated in Figure~\ref{fig:oss_pipepline} and \ref{fig:osm_pipepline}. Note that the pipeline models focus specifically on workflows strongly related to collaboration and highlight key differences between the two paradigms, rather than representing comprehensive development models applicable to all scenarios.

To assess the rationales of our proposed pipeline models as a prompt for the subsequent quantitative analysis, we conducted a targeted anonymous survey among experienced practitioners and researchers. Specifically, we distributed the survey through university research groups and technical forums. We first claimed that the pipelines focus on collaboration aspects rather than representing the comprehensive process at the beginning of the survey, and included a screening question regarding the participants' experience in OSS or OSM projects to ensure data quality. Then we included closed-ended questions asking participants to evaluate the representativeness of the OSS and OSM pipelines, with options to \textit{``agree''}, \textit{``don't agree''}, or \textit{``not sure''}. We also set an optional open-ended section for qualitative feedback at the end of the survey.

After a one-week period, we collected 179 valid responses, with 84 respondents (46.9\%) identified as software engineers and 95 respondents (53.1\%) identified as academic researchers. Notably, 136(75.9\%) of the participants reported having more than three years of experience in open collaboration projects. As for OSS Pipeline, 147 respondents (82.1\%) agreed with the proposed workflow, with 32 (17.9\%) being \textit{``not sure''}, while for OSM Pipeline, 151 respondents (84.4\%) agreed, with 28 (15.6\%) being \textit{``not sure''}. Crucially, no respondents expressed explicit disagreement. Although a few respondents provided qualitative comments suggesting more possible downstream usage approaches, which have limited correlation with collaborative development processes, the overall high approval rates suggest that our models capture the core phases of the development paradigms, providing empirical support for the rationality of the proposed pipelines.

It is important to clarify that the proposed models are intended to serve as a prompt for subsequent quantitative investigation, rather than as a universal abstraction capturing every nuance of all collaborative development scenarios. Thus, this survey does not aim to verify the absolute correctness of the models across all contexts. Instead, the obtained community consensus and face validity~\cite{nevo1985face} provide a sufficient foundation for the purposes of a preliminary analysis.

Comparing the validated pipelines of OSS and OSM development reveals different operational flows. OSS development primarily centers on code-related tasks, where downstream modifications are typically merged back into the upstream repository, forming a cyclical contribution loop. In contrast, the OSM pipeline incorporates unique upstream phases such as data engineering and model training, while its downstream usage often involves fine-tuning, where the model serves as a foundation for specialized applications rather than returning to the source.

These pipeline differences suggest three key dimensions where OSS and OSM development paradigms may exhibit measurably different collaborative patterns. First, the presence of distinct phases like model architecture construction and training in OSM, as opposed to the coding tasks in OSS, motivates a comparison of \textbf{collaboration intensity (RQ1)}. Second, the inclusion of data engineering and training stages in the OSM pipeline implies that the accessibility and transparency of resources may differ, prompting an examination of \textbf{collaboration openness (RQ2)}. Third, the downstream participation patterns suggest potential variations in \textbf{user innovation (RQ3)}, which may manifest in different \textbf{communication content} in each community. Finally, the need to move beyond statistical comparisons and interpret these findings within real-world context leads to the exploration of expert perspectives on \textbf{the potential socio-technical factors (RQ4)} shaping these collaborative paradigms.

These preliminary observations prompt subsequent empirical investigation, guiding our quantitative analysis focusing on measuring collaboration intensity, openness, and communication content, while setting the context for elucidating the potential underlying factors of these differences through expert perspectives.

\section{Methodology}\label{sec:method}

\begin{figure}[b]
    \centering
    \vspace{-.5em}
    \includegraphics[width=0.98\linewidth]{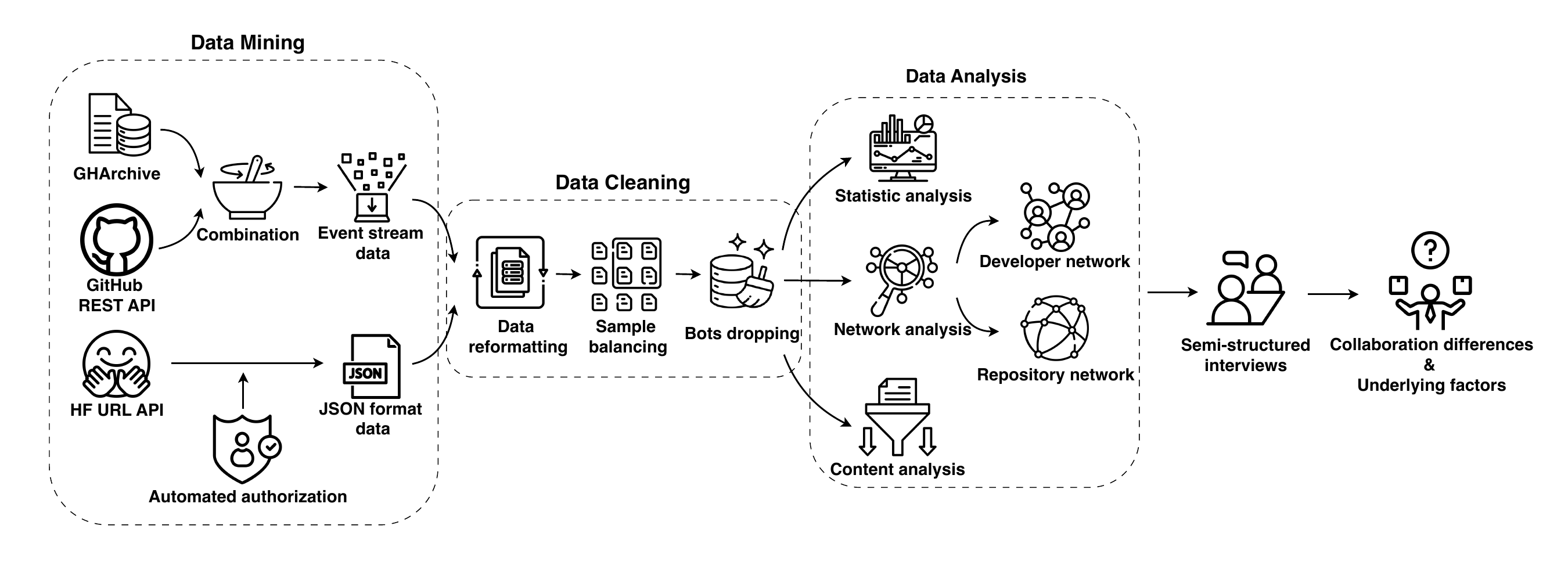}
    \vspace{-1.5em}
    \caption{Overview of methodology}
    \label{fig:method}
    \Description{methodology of the study including data mining, data cleaning, data analysis, semi-structured interviews and summarizing findings}
    \vspace{-1em}
\end{figure}

We combined a large-scale quantitative analysis and follow-up semi-structured interviews to answer the research questions, as shown in Figure \ref{fig:method}. We conducted comprehensive statistical, social network and content textual analyses of OSS and OSM ecosystems using data collected from GitHub and HF Hub, systematically measuring and characterizing differences in collaboration intensity, openness, and user innovation between OSS and OSM development paradigms. These quantitative findings were further contextualized through follow-up semi-structured interviews to uncover the socio-technical factors behind the observed differences.

\subsection{Data Collection}
\subsubsection{Data Mining}\label{subsubsec:data_mine}

To construct the dataset for empirical quantitative analysis, we first collected data from both OSS and OSM ecosystems. GitHub and HF Hub represent the main platforms with significant global popularity and activity levels for developing and sharing OSS and OSM, respectively. Given this predominant popularity and to constrain our data collection to a manageable and operationally feasible scope, we extracted data from GitHub and HF Hub for later analysis. Note that although other online hosting platforms such as GitLab and PyTorch Hub were omitted here, which may exhibit different cultures, our research focuses on dominant trends within open source development paradigms. Moreover, the widespread practice that projects originally hosted on alternative platforms create mirrors or migrate to GitHub and HF Hub~\cite{cosentino2016findings,roveda2017does} mitigates potential biases arising from our data source selection strategy, which is discussed in detail in Section \ref{sec:discuss}.

GitHub officially provides REST API~\cite{restapi} for retrieving data, and given its established role as a critical data source for OSS analysis, some well-maintained and regularly-updated GitHub datasets, such as GHArchive~\cite{gharchive}, are available for researchers as a cost-effective data acquisition approach. However, GHArchive contains only original event stream information accessible through endpoints for events without further expansion, and text content in GHArchive, such as raw text of issue descriptions, is often truncated. Therefore, we started with GHArchive to collect the event stream information from all repositories hosted on GitHub to date, and then utilized the REST API to gather and supplement communication and collaboration content including complete issue descriptions, commit messages, PR messages, etc. We accomplished this process with 437,636,881 GitHub repositories extracted.

The rapid advancement of AI has resulted in a boom of the number of models stored on HF Hub, creating a need for up-to-date data collection as existing datasets quickly become outdated. Lack of communication content and collaboration relationship in existing datasets also contribute to the need for a new updated dataset~\cite{jiang2023ptmtorrent, ait2023hfcommunity, jiang2024peatmoss}. We utilized the HF URL API to mine data from HF Hub. Considering some models on HF Hub require permission to get access, we developed an automated request script that identified models requiring authorization during the mining process and automatically requested access to ensure the completeness of data mining. Ultimately, we collected data from 1,440,527 repositories, including 1,245 models accessed after automated authorization out of 18,489 unauthorized models identified, and created the most updated and comprehensive HF Hub dataset to the best of our knowledge.

\subsubsection{Data Cleaning}

In this section, we performed data reformatting, sample balancing, and bot dropping procedures to create an operational dataset for data analysis.

\textbf{Data reformatting.} First, we restructured the collected OSS and OSM repository samples. For OSS data obtained from GHArchive and REST API, the original format consisted of event streams (i.e., chronological sequences of all events) and their associated metadata within each repository. We aggregated these event streams and transformed them into a standardized format with repositories as the primary objects. Similarly, the OSM data retrieved from the HF URL API consisted of raw JSON data with complex hierarchical structures that made searching and utilization of target elements inconvenient. Thus, we transformed the JSON-formatted data into the same repository-centric dataframe used for OSS data.

\textbf{Sample balancing.} Given that OSM emerged later than OSS and HF Hub was established in 2016, we ensured temporal alignment by excluding all GitHub OSS repositories created before 2016. Even after this adjustment, the number of OSS repositories retrieved from GitHub substantially exceeded the number of OSM repositories retrieved from HF Hub. To achieve comparable sample sizes, we performed stratified random sampling~\cite{cochran1946relative} based on star counts from the 437,636,881 GitHub repositories, resulting in a balanced sample of 1,428,792 OSS repositories that closely matched the scale of our OSM sample.

\textbf{Bots dropping.} Prior research~\cite{chidambaram2022bot, wessel2018power} has revealed the prevalence of automated bots in open-source projects that perform routine updates and integration to support development. Since these bots do not represent actual developers, their activities cannot be considered human interactions within the development process. To ensure that our research focused specifically on human interactions and eliminated interference, we implemented a bot dropping procedure. Specifically, through manual inspection and literature synthesis~\cite{chidambaram2022bot,  abdellatif2022bothunter}, we created a bot list and identified users whose names matched the entries in the list. As a result, we identified and removed 69,847 bots out of 26,030,556 users and 1,155 bots out of 642,773 users from the OSS and OSM ecosystem, respectively, allowing the datasets to more accurately reflect interactions between humans.

\subsection{Data Analysis}
\subsubsection{Statistical Analysis}\label{subsubsec:static_analysis}
To investigate the differences in collaboration intensity between OSS and OSM development processes, we first conducted statistical analyses on our collected OSS and OSM datasets.

Despite being different hosting platforms, both GitHub and HF Hub utilize \textit{git} version control system as their underlying structure, and share similarities in platform-user interaction design. This enabled us to identify comparable features that reflect equivalent concepts across both platforms. Considering that, in knowledge sharing and collaboration, there are interactions of varying intensity~\cite{qin1997types}, which also applies to the open-source development process, we selected features from both GitHub and HF Hub repositories that represent different levels of communication and collaboration from highest to lowest intensity.

For OSS projects hosted on GitHub, we analyzed commits, issues (including their replies and discussions), and stars, while for OSM projects hosted on HF Hub, we analyzed commits, discussions, and likes. Commits represent direct contribution to a repository, and multiple developers committing to the same repository indicates collaboration and communication of relatively high degree~\cite{dabbish2012social}; issues and discussions contain topics including feature suggestions, bug reports, and user problems and reflect developers' understanding of a repository and direct communication on it, signifying moderate collaboration but strong knowledge exchange~\cite{tsay2014let}; stars and likes show developers' preferences for repositories~\cite{borges2018s}, and as noted by Fang et al.~\cite{fang2024strength}, connections formed between developers and repositories through starring constitute ``weak ties'' that indicate lower-level communication and knowledge sharing between developers and repositories respectively.

To validate that selected features represent different aspects of communication and collaboration, we calculated Pearson correlation coefficients of these features using all the data in our OSS and OSM datasets. Subsequently, we performed a large-scale statistical analysis on the complete datasets to characterize the distribution patterns of these three features across both datasets, including calculating basic statistical measures to illustrate differences in the collaboration intensity between OSS and OSM development processes. Furthermore, we applied the Mann-Whitney U test~\cite{nachar2008mann} to selected features across OSS and OSM datasets to demonstrate the statistical significance of the observed differences.

\subsubsection{Social Network Analysis}

To further investigate the difference in collaboration intensity and openness between OSS and OSM development processes, we constructed and analyzed social networks based on the preceding features in a subset of the collected dataset. We describe the construction method as follows and the analysis results are illustrated in Section \ref{sec:result}.

\textbf{Repository Network.} We constructed repository networks by regarding repositories as nodes. When the same developer performed interactions of the same type on two different repositories within a specific time period, we established an edge between the repository nodes. We assume that connections built between repositories due to the actions of developers symbolize a form of communication and knowledge exchange between these repositories, further reflecting patterns of communication and knowledge exchange within the open-source development ecosystem~\cite{sowe2008understanding, fang2024strength}.

As mentioned earlier, commits, issues/discussions, and stars/likes represent indicators of knowledge exchange in decreasing order of intensity. Among these, stars/likes primarily show preferences of stargazers, representing low-level interaction. We aim to consider communication and collaboration of relatively higher intensity, so we chose to construct four repository networks (two for each dataset) based on commit and issue/discussion activities, corresponding to direct contribution and knowledge exchange in OSS and OSM development contexts, respectively. 

Specifically, we first selected the top 10,000 repositories by star count and like count from the OSS and OSM dataset as subjects for following network analysis, respectively. We conducted this process for two purposes: (1) As detailed in Section \ref{sec:result}, both OSS and OSM projects exhibit skewed distributions in activity levels, particularly OSM projects. Selecting repositories with higher star/like counts effectively filters out trivial ones, allowing our research to focus on active projects that better reflect communication and collaboration states in the development process; (2) We took computational burdens into consideration, as analyzing the entire collected dataset would create excessively large networks that would pose significant computational challenges while being less likely to change experimental results. Subsequently, we checked commit and issue/discussion operations within the 10,000 repositories for both OSS and OSM, identifying the developers who performed the operations and their occurrence timestamps. Considering that operations performed by the same developer but separated by long time intervals may not effectively represent communication and knowledge exchange, we adopted a one-year time window, which was also employed by Fang et al.~\cite{fang2024strength}. We recorded an undirected edge between two repositories if and only if both were interacted with by the same developer within the one-year time window.

\textbf{Developer Network.} While repository networks illuminate one aspect of communication and knowledge exchange patterns, developer networks, which designate individual developers as nodes, reflect collaboration patterns among developers in open-source development processes more directly.

As for developer network construction, we considered commits that represent indicators of high-intensity collaboration and issues/discussions that represent knowledge exchange based on project understanding, excluding stars/likes because they merely indicate preferences rather than substantive collaboration from the developer's perspective~\cite{borges2018s}. And we did not apply time window restrictions when constructing developer networks, as active human-to-human collaboration commonly accommodates temporal delays in open-source development, and time differences between collaborative operations do not significantly change the collaborative nature of these interactions~\cite{herbsleb2000distance, gutwin2004revealing}. Consequently, we recorded an undirected edge between two developers if and only if they interacted with the same project repository, constructing developer networks using the 10,000 OSS repositories and 10,000 OSM repositories.

\subsubsection{Content Analysis}\label{subsec:content}

User innovation in OSS development is driven by contributors who contribute their expertise and requirements to the software~\cite{hippel2003open, bogers2012managing}. This process is often evidenced by the messages they leave in repositories and community~\cite{treude2011effective}. By analyzing these messages, we can trace potential divergence in user innovation between OSS and OSM development.

Among the features we selected, commit messages and issue/discussion content could potentially reflect the communication and collaboration content in development processes. However, considering that in development practice, commit messages are often missing or trivial (i.e., directly using the template without useful information)~\cite{tian2022makes}, we examined the collected data and found significant flaws with commit messages. Our inspection revealed that within the 10,000 GitHub repositories in the OSS subset, which contained 23,632,337 commit records, only 817,679 (3.46\%) included substantive commit messages after excluding blank and trivial ones. Similarly, in the OSM subset of 10,000 HF Hub repositories containing 212,085 commit records, only 11,706 (5.52\%) included meaningful commit messages. We assumed that substantial missing data would introduce bias into a commit message-based content analysis because results would disproportionately represent developers who habitually write detailed commit messages rather than common developers in the broader community, thus failing to accurately reflect collaboration and communication patterns in the open-source development processes.

In contrast, issues/discussions represent active interactions initiated by developers, in which the content itself constitutes the developer's intention and contribution, ensuring the meaningfulness of most issue/discussion content. Furthermore, as channels for knowledge exchange based on developers' understanding of their own or others' projects, issues/discussions more directly reflect the thematic content of collaboration and communication in open-source development processes, further reflecting how user innovation works. Therefore, we finally chose to conduct thematic content analysis on issue/discussion texts.

As LLMs have demonstrated
superior capability in NLP tasks such as text annotation~\cite{gilardi2023chatgpt, wang2025lata}, we employed an AI-assisted thematic analysis approach to conduct the investigation. In contrast to traditional topic modeling methods like Latent Dirichlet Allocation (LDA)~\cite{jacobi2018quantitative}, which rely on statistical word co-occurrences and may overlook nuanced, context-dependent semantics, our approach leverages LLMs' strength in natural language understanding to extract contextualized themes from unstructured text. Moreover, to address concerns regarding the black-box nature of LLMs and potential hallucinations, we implemented a multi-stage, human-in-loop workflow combining Chain-of-Thought (CoT)~\cite{wei2022chain} processing with human verification~\cite{braun2006using} as follows.

\textbf{Automated Initial Coding with CoT}. We utilized \texttt{Claude 3.7 Sonnet} and designed a CoT prompting strategy to improve the interpretability and rigor of this step, which instructed the LLM to explicitly output intermediate reasoning steps: summarizing the discussion context, extracting keywords, and proposing a preliminary label. This step engaged the LLM in data familiarization and initial code generation, transforming it from an black-box predictor into a transparent analytical assistant with traceable reasoning.

\textbf{Human Verification and Reliability Check}. To validate the reliability of LLM-generated labels, two authors independently performed a manual labeling process on 100 randomly selected samples. Note that, for messages with mixed intents, we assigned the label based on the primary resolution provided by maintainers. First, to ensure the reliability of human baseline, the two authors discussed and resolved all discrepancies in their initial open codes, thereby establishing a unified human-annotated ground truth. Then we compared the consensus results with the LLM's outputs. Since the initial open coding process was inductive, direct statistical comparison was not applicable due to inconsistent codes. Therefore, we performed a post-hoc semantic alignment~\cite{campbell2013coding}, systematically mapping the varied descriptors from both LLM and human ground truth to a standardized taxonomy based on their underlying definitions. For instance, synonymous terms \textit{``login error''} labeled by LLM and \textit{``auth fail''} labeled by human were normalized under the unified category \textit{``Authentication Issue''}. After normalization, we found that the semantic concepts were consistent across the samples, with no hallucinations observed.

\textbf{Theme Construction and Refinement.} Moving from codes to themes was a human-led process. We first carefully reviewed the codes and summaries generated by LLM. Then we iteratively grouped codes into broader themes (e.g., \textit{``Unexpected Crashes''} and \textit{``Unresponsiveness''} $\rightarrow$ \textit{``Application Instability''}; \textit{``Application Instability''} and \textit{``Authentication Issue''} $\rightarrow$ \textit{``Bug Report''}), reviewed them against original data for consistency, and defined the final thematic map. To ensure conciseness and facilitate comparison between OSS and OSM, we ranked the finalized themes by their prevalence. We applied a cumulative coverage threshold, retaining themes accounting for the top 90\% of the samples as distinct categories, while aggregating the remaining long-tail themes into \textit{``Others''}.

\subsection{Semi-Structured Interview}

To triangulate our quantitative findings and explore the socio-technical factors underlying the observed differences between OSS and OSM development paradigms, we conducted a semi-structured interview to collect qualitative opinions from experts that contextualize the identified quantitative findings.

\subsubsection{Participants and Recruitment}

We employed a purposive sampling strategy to recruit practitioners who possessed dual-domain expertise in both OSS development and OSM development. This criterion ensured that participants could offer comparative insights based on firsthand experience in both development ecosystems.

Instead of pre-determining the sample size, we adopted a sequential recruitment strategy based on the principle of thematic saturation~\cite{guest2006many, saunders2018saturation}. We recruited and interviewed participants iteratively, analyzing data concurrently with data collection. We ceased recruitment after the 10th participant (P1-P10, see Table \ref{tab:interviewee}), at which point we observed that the explanatory themes regarding the socio-technical factors underlying the OSS and OSM development had stabilized. The final group of participants consisted of experts from diverse backgrounds including leading tech companies (e.g., Intel, Microsoft, etc.), academic research labs, and independent open-source communities.

\begin{table}[t]
	\centering
	\small
	\caption{List of interviewees}
    \vspace{-.5em}
	\begin{tabular}{p{0.5cm}p{3.3cm}p{5.3cm}p{3.2cm}}
	\toprule
    	\textbf{ID} & \textbf{Sector} & \textbf{Main Role} & \textbf{Experience (OSS/OSM)} \\ \midrule
    	P1 & Tech Company (IT) & Model Producer, Code Contributor & 8 years / 5 years \\
    	P2 & Tech Company (IT) & Community Manager, Code Contributor & 10 years / 4 years \\
    	P3 & Academic Research Lab & Academic Researcher & 6 years / 4 years \\
    	  P4 & Tech Company (IT) & Model
          Producer, Code Contributor & 10 years / 4 years \\
    	P5 & Academic Research Lab & Academic Researcher & 5 years / 5 years \\
    	P6 & Independent Community  & Code Contributor & 4 years /  3 years \\
    	P7 & Tech Company (IT) & Model Producer, Code Contributor & 6 years / 4 years \\
    	P8 & Tech Company (IT) & Community Manager & 8 years / 3 years \\
        P9 & Tech Company (IT) & Community Manager, Code Contributor & 6 years / 3 years \\
        P10 & Independent Community & Code Contributor & 3 years / 2 years \\
    \bottomrule
	\end{tabular}
    \Description{Ten interviewees included in the interview process}
	\label{tab:interviewee}
\end{table}

\subsubsection{Interview Process}

We conducted 10 semi-structured online interviews, and each session lasted
approximately 60 minutes. We first asked participants to describe their experience in collaborative development of OSS and OSM projects. Subsequently, we presented participants with the quantitative key findings and corresponding visualizations~\cite{tubaro2016visual, kallio2016systematic}, asking them to reflect on whether these findings aligned with their practical experience and observation. Then we elicited participants' responses based on preceding findings following a semi-structured guide with
four topics: their workflows
and collaboration habits in OSS and OSM projects; their opinions on the factors that drive the differences of collaboration between OSS and OSM development; their opinions on innovation patterns in OSS and OSM development; and their opinions on approaches to contribute the collaboration and innovation in OSM development.

\subsubsection{Analytic Procedure}

We followed a reflexive thematic analysis approach to analyze the interview data~\cite{braun2019reflecting, braun2006using}. The process followed three steps: inductive open coding, theme development, and consensus building and member checking. 
In particular, two authors first independently performed line-by-line coding on the transcripts to create codes that capture the specific socio-technical factors described by participants. Two authors then iteratively engaged in sorting and collating codes into broader themes. These themes were synthesized to construct coherent narratives that explain the observed development paradigm divergence. Finally, two authors resolved initial discrepancies through iterative consensus meetings, and conducted member checking~\cite{varpio2017shedding} by sharing the finalized themes with participants to ensure accuracy and practical relevance.

\section{Results}\label{sec:result}

\subsection{RQ1: Collaboration intensity}

\begin{figure}[thbp]
	\centering
        \vspace{-.5em}
	\begin{minipage}{0.45\linewidth}
		\centering
		\includegraphics[width=0.9\linewidth]{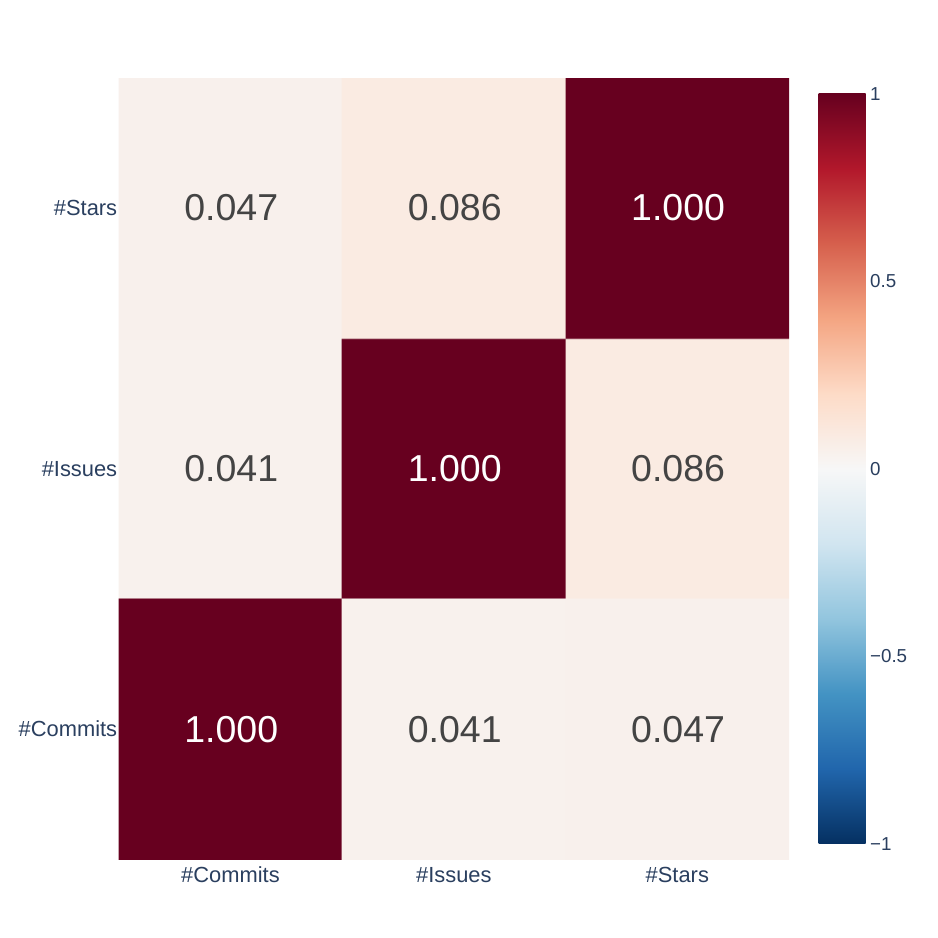}
		\caption{Correlations of selected features in OSS}
        \Description{Num of commits, issues and stars has low correlation with one another}
		\label{fig:oss_correlation}
	\end{minipage}
	\qquad
	\begin{minipage}{0.45\linewidth}
		\centering
		\includegraphics[width=0.9\linewidth]{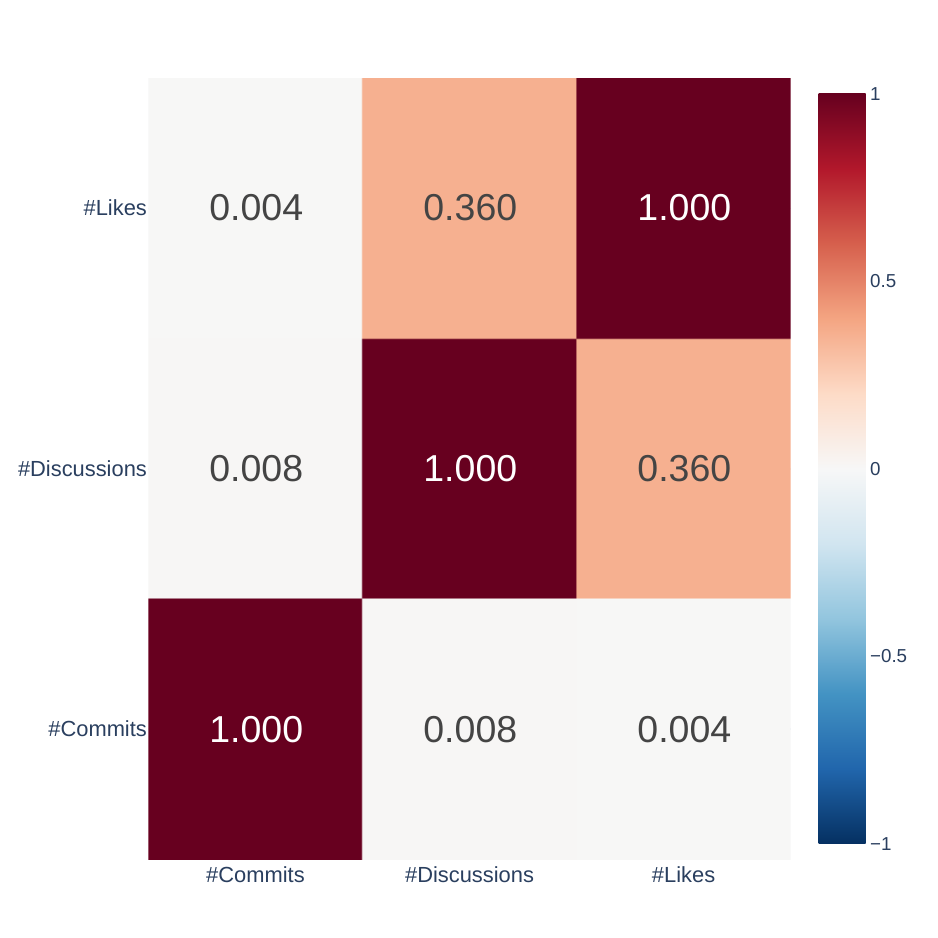}
		\caption{Correlations of selected features in OSM}
        \Description{Num of commits, discussions and likes has low correlation with one another}
		\label{fig:osm_correlation}	\end{minipage}
\end{figure}

The Pearson correlation analysis of commit, issue/discussion, and star/like parameters in both OSS and OSM datasets is presented in Figure \ref{fig:oss_correlation} and \ref{fig:osm_correlation}. The results demonstrate positive but weak correlations between these parameters in both ecosystems, validating our assumption in Section \ref{sec:method} that these parameters represent different aspects of developer communication and collaboration within OSS and OSM development ecosystems.

\begin{figure}[bhtp]
    \centering
    \begin{subfigure}{0.7\textwidth}
        \centering
        \includegraphics[width=\textwidth]{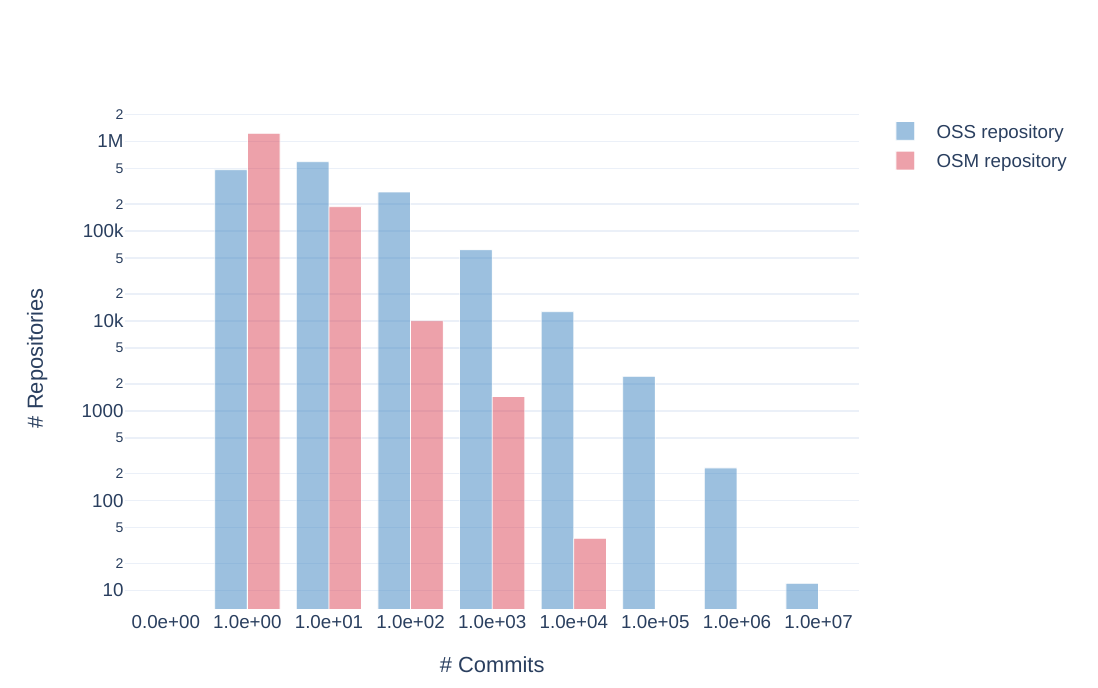}
        \caption{Distribution of commits of OSS and OSM repositories}
        \Description{Distribution of commits of OSS exceeds that of OSM}
        \label{subfig:commit_dis}
    \end{subfigure}
    \hfill
    \begin{subfigure}{0.7\textwidth}
        \centering
        \includegraphics[width=\textwidth]{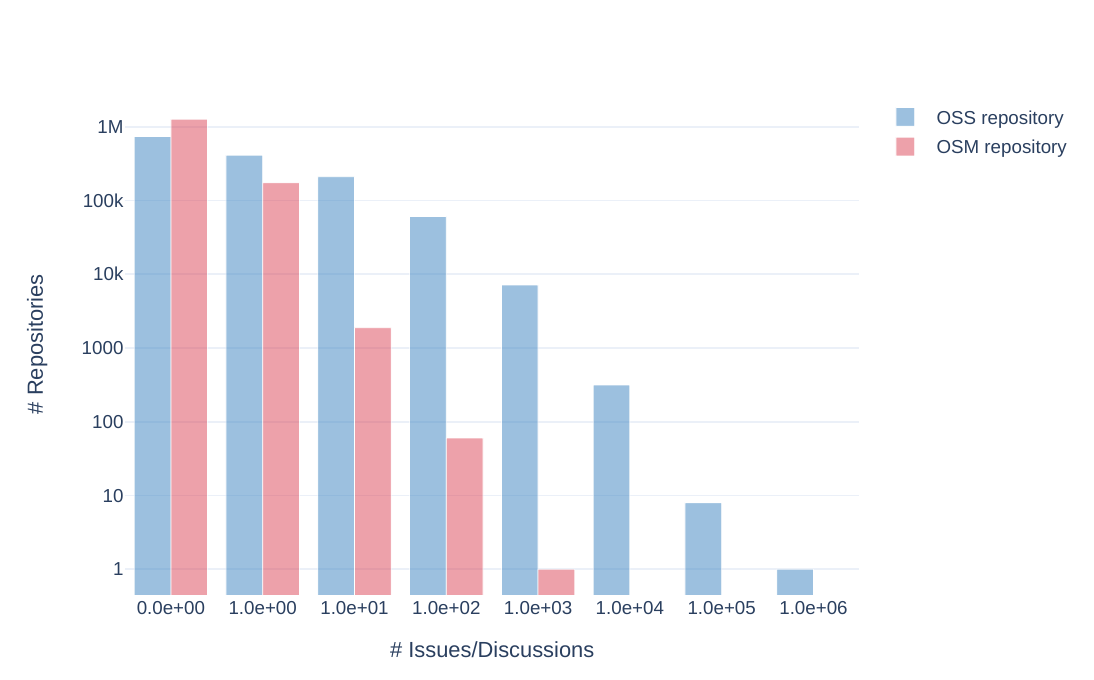}
        \caption{Distribution of issue/discussion of OSS and OSM repositories}
        \Description{Distribution of issues of OSS exceeds that of OSM}
        \label{subfig:issue_discussion_dis}
    \end{subfigure}
    \hfill
    \begin{subfigure}{0.7\textwidth}
        \centering
        \includegraphics[width=\textwidth]{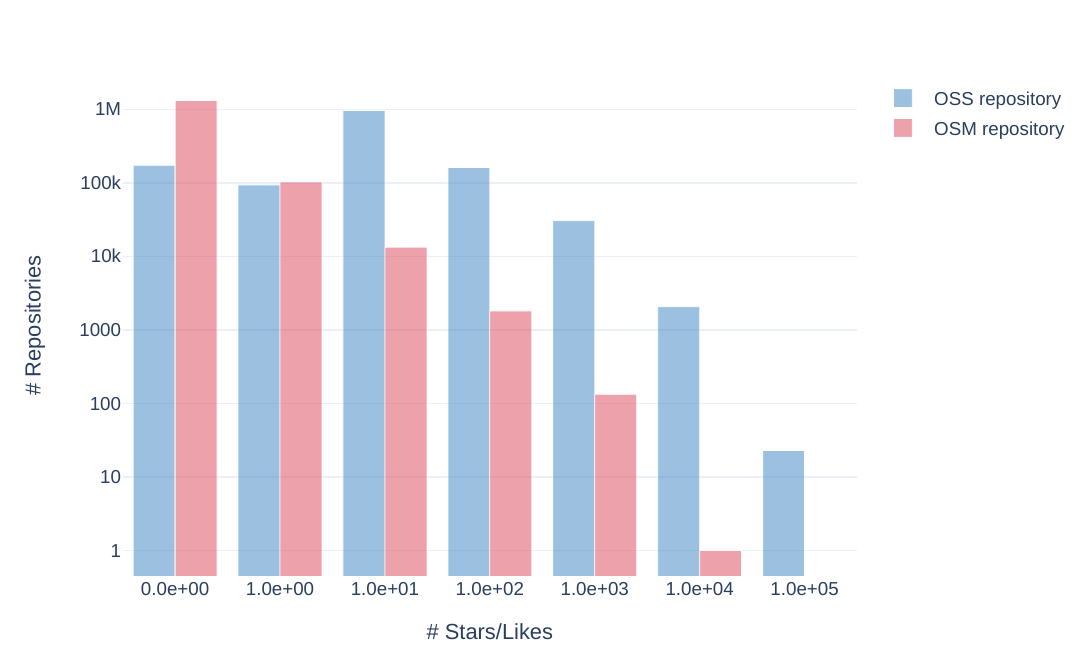}
        \caption{Distribution of star/like of OSS and OSM repositories}
        \Description{Distribution of stars of OSS exceeds that of OSM}
        \label{subfig:star_like_dis}
    \end{subfigure}
    \vspace{-.5em}
    \caption{Distribution of Activities of OSS and OSM repositories}
    \label{fig:OSS_OSM_dis}
\end{figure}

Figure~\ref{fig:OSS_OSM_dis} illustrate the distribution of these parameters across the OSS and OSM datasets, respectively. Both datasets exhibit skewed distributions, with this pattern particularly pronounced in OSM dataset, where the vast majority of repositories contain extremely few commits and discussions, while a small minority contain substantial commits and discussions. Comparing the distributions between OSS and OSM reveals two key insights: (1) OSS community activities and collaboration intensity generally exceed those observed in OSM; (2) parameter distributions in OSS demonstrate less skewness than in OSM, indicating that collaboration occurs more frequently and is more uniformly distributed across the OSS ecosystem, rather than being concentrated within a few prominent projects.

\begin{table}[t]
  \caption{Statistical analysis results of OSS and OSM repositores}
  \label{tab:statistical}
  \begin{tabular}{cccc}
    \toprule
    Attribute&Mean&Median&Max\\
    \midrule
    \#Commits (OSS) & 1464.28 & 24.00 & 71860646.00\\ 
    \#Commits (OSM) & 10.65 & 2.00 & 65092.00\\
    \midrule
    \#Issues (OSS) & 35.94 & 0.00 & 1607850.00\\
    \#Discussions (OSM) & 0.20 & 0.00 & 3010.00\\
    \midrule
    \#Stars (OSS) & 142.78 & 16.00 & 318751.00\\
    \#Likes (OSM) & 0.93 & 0.00 & 10194\\
  \bottomrule
\end{tabular}
\Description{Three attributes of OSS all exceed that of OSM}
\end{table}

Table \ref{tab:statistical} presents the results of statistical analysis for the three parameters in both datasets. Accordingly, OSS repositories exhibit significantly higher activity levels compared to OSM repositories across all metrics, with the magnitude of activity differing by more than 100-fold. And the Mann-Whitney U tests further provide strong evidence of statistically significant differences between selected parameters in OSS and OSM repositories with all tests having p < 0.001. It indicates from a macroscopic perspective that the intensity of collaboration in OSM ecosystem lags behind that in OSS ecosystem in development contributions, project understanding and knowledge exchange, and preference expression.

Following the methodology described in Section \ref{sec:method}, we constructed repository networks and user networks based on commit and issue/discussion connections and conducted network analyses, with results presented in Table \ref{tab:network}.

For repository networks, OSS exhibits a significantly lower node connection rate in commit networks compared to OSM, along with lower average node out-degree. Although this initially appears counterintuitive, it aligns with logical expectations when properly contextualized. It is important to recall that our network analysis focused on the top 10,000 projects by star/like count, effectively filtering out trivial projects, and in repository commit networks, connections between repositories are established if and only if the same developer contributes to both repositories. Therefore, lower repository node connection rate indicates that among nontrivial OSS repositories, developer composition is more diverse rather than being dominated by the same group of a few developers. Conversely, the higher node connection rate and out-degree in OSM commit networks suggest that most high-impact projects are developed by the same cohort of developers. Issue/discussion networks emphasize project understanding and knowledge exchange. The repository issue network of OSS demonstrates a higher node connection rate and average out-degree compared to OSM. This indicates that despite the more diverse developer composition across OSS projects, communication between projects occurs more frequently, with developers more willing to understand and engage with multiple different projects, actively participating in knowledge exchange.

For user networks, OSS demonstrates a higher node connection rate and average node out-degree in both commit networks and issue networks compared to OSM. This reflects that developers in the OSS ecosystem engage in more extensive collaborations than those in the OSM ecosystem, both in terms of cooperative development and project understanding with knowledge exchange.

In summary, the intensity of collaboration in OSM development paradigm is significantly lower than that in OSS development paradigm, with measurable metrics revealing 100-fold differences, significant Mann-Whitney U test results, and substantial disparities in social network analysis.

\begin{table}[th]
\caption{Network analysis result of OSS and OSM repositories}
\vspace{-.5em}
\label{tab:network}
\begin{tabular}{cccccc}
\toprule
\makecell[c]{Network\\type} &\makecell[c]{Attribute}& \makecell[c] {Connected\\rate}  &\makecell[c]{Average\\degree (all)} &\makecell[c]{Average\\degree (connected)} &\makecell[c]{Max\\degree} \\ 
\midrule
\multirowcell{4}{Repository}&Commit (OSS)&47.80\%&1.44&3.01&31\\
                   &Commit (OSM)&80.07\%&68.06&85.00&478\\
                   &Issue (OSS)&99.15\%&328.92&331.74&1353\\
                   &Discussion (OSM)&85.34\%&135.67&158.97&1395\\
\midrule
\multirowcell{4}{User}&Commit (OSS)&98.37\%&422.20&429.19&1914\\
                   &Commit (OSM)&92.04\%&7.27&7.90&367\\
                   &Issue (OSS)&100\%&5880.94&5880.94&93441\\
                   &Discussion (OSM)&99.96\%&139.45&139.50&7701\\
\bottomrule
\end{tabular}
\Description{Network analysis result shows that OSS has higher collaboration intensity and openness than OSM}
\end{table}

\subsection{RQ2: Collaboration openness}\label{subsec:rq2}

The preceding analysis of repository commit network node connection rates and average node out-degrees preliminarily indicates that the OSS development ecosystem features more diverse developer composition, with prominent projects not monopolized by the same group of developers. This indirectly suggests that collaboration among developers in the OSS ecosystem may extend across broader scopes and exhibit higher levels of openness than OSM.

To investigate this phenomenon more directly, we selected repositories from our OSM dataset with organizational publishers and conducted affiliation analyses of both commit and discussion participants. We found that 79.55\% of committers belong to the publishing organization and 19.36\% are HF staff, while only 9.01\% of discussion participants are affiliated with the publishing organization. Due to GitHub's platform policies, which restrict access to comprehensive organizational membership information (allowing visibility only to publicly displayed members), we could not perform an identical analysis for OSS. However, prior research by Dias et al.~\cite{dias2018drives} has found that contributions from external developers dominate even in company-owned OSS projects with 56.7\% commits contributed. The stark contrast between these findings reveals that collaboration openness regarding direct contributions in OSM development processes is notably lower than in OSS development processes, with greater concentration within the leading companies and organizations. However, regarding communication represented by discussions, OSM development processes maintain relative openness, with external users and participants still constituting a significant proportion, which is consistent with OSS development paradigm.

\begin{figure}[hbtp]
    \centering
    \vspace{-1em}
    \includegraphics[width=0.6\linewidth]{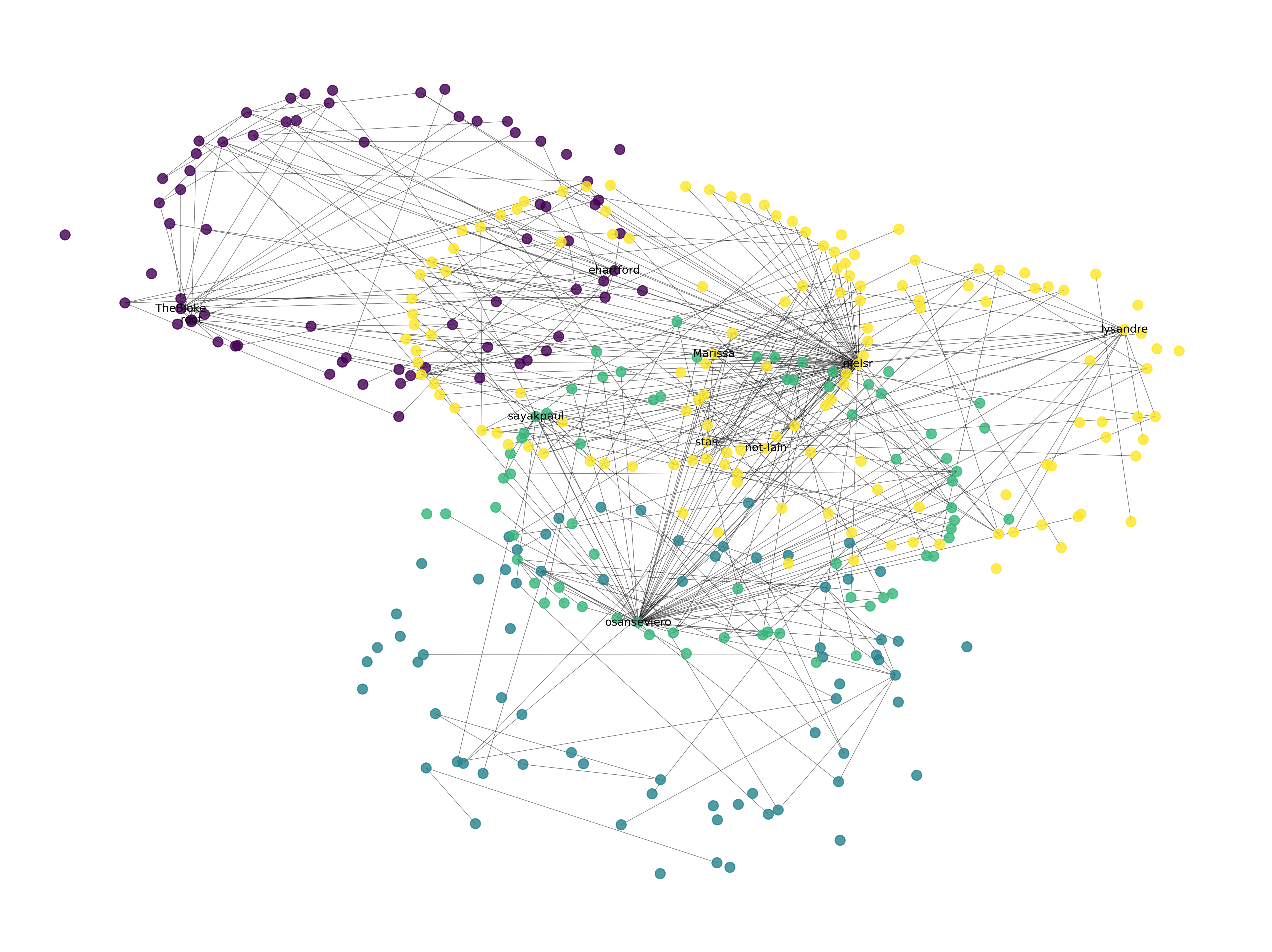}
    \vspace{-1em}
    \caption{Part of the OSM user commit network}
    \label{fig:network}
    \Description{The partial commit network and its communities}
\end{figure}

To further substantiate this finding from a macroscopic perspective, we applied the Louvain community detection algorithm~\cite{de2011generalized} to the established OSM user networks, as illustrated in Figure \ref{fig:network}\footnote{For readability, the figure displays only a small part of the network to prevent excessive node and edge density that would impair visual interpretation.}. We documented all publishing organizations corresponding to the OSM repositories used in our network analysis and obtained their membership information via the HF Hub API. We then used these developer-organization affiliations as ground truth to evaluate the overlap with communities identified by the Louvain algorithm.

For the user commit network, we examined the proportion of community members belonging to the predominant organizational affiliation within each detected community across the entire network. This proportion averaged 49.8\% and notably, some actual organizations do not have corresponding organization accounts on HF Hub, making them unobservable in this analysis. Therefore, the actual proportion would be even higher. This result reflects consistency between contribution-based collaboration communities and actual organizational affiliations. Meanwhile, the same metric for the user discussion network reached only 9.2\%, further demonstrating that project understanding and knowledge exchange represented by discussions in the OSM ecosystem maintain relatively openness despite the more closed collaboration patterns in development contributions.

In summary, the openness of collaboration regarding direct contribution in OSM development paradigm is substantially
lower than that in OSS development paradigm, with organizational model releases showing only 1.09\% external contribution rates and contribution-based community detection revealing over 49.8\% alignment with actual organizational affiliations. However, OSM maintains more open knowledge exchange despite relatively closed development practices, with discussion-based community detection demonstrating only 9.2\% alignment with actual organizational affiliations.

\subsection{RQ3: User innovation}

As stated in Section \ref{sec:method}, manifestation of user innovation in open source development is reflected by communication messages in repositories and community, thus we analyze these messages to  trace potential divergence in user innovation.

\begin{figure}[htbp]
    \centering
    \includegraphics[width=0.7\linewidth]{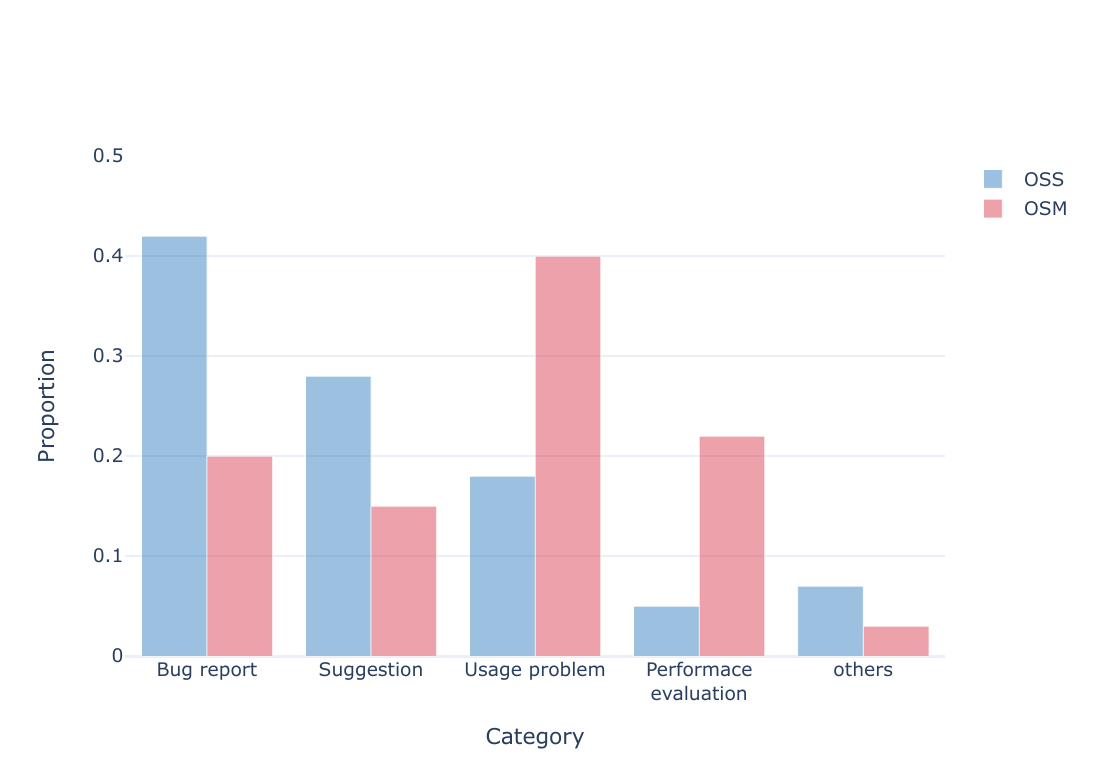}
    \vspace{-1em}
    \caption{Distribution of communication theme of OSS and OSM repositories}
    \label{fig:content_theme}
   
    \Description{The proportion of the communication content theme of OSS and OSM}
    \vspace{-.5em}
\end{figure}

Following the AI-assisted and human-in-loop approach described in Section \ref{sec:method}, we employed LLM and conducted a thematic analysis, with results presented in Figure \ref{fig:content_theme}. We categorized communication content into five thematic categories: \textit{bug reports} (reporting unintended system defects or failures), \textit{suggestions} (proposing capability extensions for correctly functioning artifacts), \textit{usage problems} (discussing local configuration issues or knowledge gaps, i.e., ``how-to'' questions), \textit{performance evaluation} (sharing benchmarking results or experiential feedback), and \textit{others} (aggregating the remaining long-tail themes). The result demonstrates that, in OSS contexts, \textit{bug reports} constitute the predominant content category (42.7\%), followed by \textit{suggestions} (28.2\%). This distribution indicates that OSS communication substantially focuses on identifying and resolving problems within the codebase and enhancing functionality through collaborative improvement. However, for OSM, \textit{usage problems} represent the largest content category (40.0\%), and the second most prevalent category is \textit{performance evaluation} (22.3\%). This distribution suggests that OSM communication is primarily usage-oriented and less related to the development process.

These findings empirically quantify and characterize the differences in communication content between OSS and OSM development paradigms. OSS communication content encompasses more contribution-focused discussions aimed at improving the project through collaborative problem-solving and feature enhancement. OSM communication content, by contrast, centers on usage-centered discussions about model utilization and feedback, with less emphasis on collaborative improvement of the underlying model itself. 

More fundamentally, the markedly different communication content between OSS and OSM ecosystems reflects distinct user innovation patterns. To be specific, OSS communication dominated by bug reports and improvement suggestions demonstrates \textbf{collaborative improvement innovation} around the software, where users actively participate in enhancing collaborative artifacts through direct modification and collective problem solving~\cite{hippel2003open, bogers2012managing}. Whereas OSM communication dominated by usage-centered content reflects \textbf{adaptive utilization innovation} around the model, where users focus on discovering novel applications and optimizing deployment through downstream adaptations (e.g., fine-tuning), rather than contributing iterative improvements to the base models~\cite{jiang2023empirical, wortsman2022robust, ziegler2019fine}.

Within the OSS development paradigm of collaborative improvement innovation, users engage as co-developers in continuous artifact enhancement, meanwhile benefiting from improved products resulting from these collaborative contributions~\cite{hippel2003open, von2007open}. In contrast, adaptive utilization innovation within the OSM development paradigm represents a distinct user innovation pattern occurring in downstream application space rather than base model space. Users position themselves as adaptive implementers that create value through novel utilization and domain-specific integration~\cite{wortsman2022robust, ziegler2019fine, liu2024optimizing}, regardless of whether the underlying model is modified. 
However, this process still generates significant innovation value~\cite{bogers2012managing}, as it facilitates downstream knowledge generation and transfer. By sharing solutions to usage problems and results of model evaluation, users transform individual experience into public goods that illuminate model boundaries and optimal strategies. For instance, one user in HF community shared the observation when using \texttt{Llama-3.1-8B-Instruct}, stating that ``the raw prompt [a single string of system and user input] is performing better than the formatted one [using chat templates]''. Furthermore, it also drives recombinant innovation, where value is generated by integrating models into new systems or combining them for cross-domain applications. As one user described in HF discussion, ``I [connected] the model to [our] local Excel spreadsheets \dots [building] a custom sentiment analysis tool.'' Semi-structured interviews further uncovered broader insights on user innovation, which are discussed in detail in Section \ref{sec:discuss}.

In summary, the primary user innovation patterns in OSS and OSM development paradigms diverge significantly: while OSS centers on collaborative artifact improvement, OSM trends toward adaptive utilization, as characterized by different communication content within each community.

\subsection{RQ4: Underlying Factors} \label{subsec:qualitative}

Semi-structured interviews triangulated the preceding quantitative results. In particular, participants explicitly expressed that these findings align with their firsthand practical experiences, corroborating that, compared to OSS development paradigm, OSM development paradigm exhibits lower collaboration intensity, lower collaboration openness, and adaptive utilization innovation rather than collaborative improvement innovation. 

More importantly, participants collectively agreed that the divergence in collaboration paradigms between OSS and OSM is not incidental or merely a transient phase of community development, but driven by a confluence of underlying socio-technical factors. These insights are significant as they challenge the prior theory of commons-based peer production and provide a theoretical foundation for designing future AI-specific collaborative infrastructures.

\subsubsection{Technical and Architectural Barriers}

Direct collaborative improvements to OSM are significantly limited, as the non-decomposable and stochastic nature of model inhibits the modular and logic-driven participation typical in OSS.

\textit{\textbf{Loss of Traceability.}} Unlike text-based source code where logic is traceable through line-based updates, model weights are saved as binary objects that lack line-by-line differentiation. As P1 noted, ``Software code has a clear logic \dots [Model weights] are more like a data file \dots a black box without [human-readable] line-update characteristics.'' And P4 asserted that ``[Y]our changes must be traceable \dots if your changes are not traceable, how can you talk about co-working?'' Loss of traceability inhibits the comparative analysis of model iterations, effectively marginalizing external contributors who rely on human-readable semantic to verify and refine artifacts.

\textit{\textbf{Structural Non-decomposability.}} Software systems are modular, allowing distributed teams to decouple development~\cite{mockus2002two, hippel2003open}. However, the tight coupling of model structure prevents a modular division of labor across different layers. As P3 noted, ``[Y]ou cannot assign specific layers of a transformer to independent groups.'' Thus, OSM development diverges from the decentralized \textit{``bazaar''} model and tends to concentrate within centralized and specialized teams.

\textit{\textbf{Irreproducibility and Stochasticity.}} The sensitivity of model training to environmental factors and random seeds undermines reproducibility. Since reproducibility is a prerequisite for incremental development, the inability to replicate training states prevents external contributors from verifying their improvements, thereby disrupting the feedback loop essential for co-working. P2 noted that ``[I]t is nearly impossible to [perfectly] replicate the training state,'' and P4 complemented that ``The environment is more complex \dots If I want a colleague to reproduce my results [and] then make incremental improvement, I must control numerous environmental factors [that seem impossible], or they will get different outcomes.''

\subsubsection{Resource Constraints}

Collective improvements to OSM are significantly restricted by high computational resource cost and data inaccessibility.

\textit{\textbf{Computational Resource Barriers.}} Unlike traditional software development, which has minimal computational cost, validating or training AI models is financially and computationally intensive. This ``compute wall''(P8) prevents individual contributors who cannot access large-scale computing resources from collaborating on the improvement of the model itself, shifting their role to downstream users. For example, P8 recalled that ``[Our company] invested over 500 million USD last year, with 70\% spent on [computation] \dots Individual contributors simply cannot afford such participation.'' Consequently, ``the necessity for massive GPU clusters fosters a resource-driven centralization [that] relegates independent developers to the downstream,'' as noted by P5.

\textit{\textbf{Data Inaccessibility.}} The divergence from the
sharing of building logic within source code in OSS to the disclosure of training results in OSM often lead to the omission of training
data. High-quality training data is increasingly treated as a proprietary asset, shielded by privacy regulations and competitive strategies~\cite{solaiman2019release, widder2024open}. Even when companies leverage external crowd-sourcing for initial data acquisition, the subsequent ``data flywheel'' process remains strictly internal, as pointed out by P9. Since data is the foundation of model training, data inaccessibility effectively obstructs collaborative development, preventing external contributors from participating in the core development of the model.

\subsubsection{Infrastructural and Tooling Misalignment}

Existing online hosting platforms and version-control systems, originally designed for code-based artifacts, face a growing misalignment as the artifacts in OSM diverge toward complex model files. Thus, they fail to meet the functional requirements of the collaborative development of AI models.

\textit{\textbf{Platform Orientation and Immaturity.}} ``While GitHub serves as a    workshop for collaborative production, [platforms] like HF act primarily as registries for distributing [finished models],'' noted by P6. This orientation of distribution rather than collective development limits distributed co-development. Furthermore, HF has not yet provided ``necessary features for collective model building''(P10), which reflects a lack of platform maturity in supporting full development lifecycle of OSM.

\textit{\textbf{Ineffective Versioning Support.}} Standard version control tools are ineffective for managing large binary weights. Inability to track semantic updates within parameters makes collaborative efforts remain at the metadata level, preventing direct collective refinement of the model artifacts. As P7 noted, ``[W]ithout effective version control, we cannot reach the level of granularity [required] for truly collaborative development.''

\subsubsection{Corporate Strategic Factors}

In OSM, open source is often a strategic choice driven by competitive goals rather than merely a commitment to community values, which imposes limits on collaboration.

\textit{\textbf{Strategic Openness in Competitive Markets.}} Open-sourcing AI models often serves as a strategic instrument for market competition and ecosystem development, rather than a purely community-driven initiative. As P7 noted, ``[Despite of open source, ] there is hardly any collaboration between [big] companies because of commercial competition and monopoly.'' P9 complemented from another perspective that ``Startups are unlikely to open source their models because they sell [model] API as main product. Big companies have other [additive] services to sell, but for small companies, the model itself is all they have.''

\textit{\textbf{Credit Assignment Challenges.}} The complexity of assigning credit across compute, data, and algorithms makes it hard to measure isolate contribution in OSM. As P5 pointed out, ``Unlike software [where] credits can be measured by LoC [lines of code] \dots challenges in assigning credit make it [nearly] impossible to [negotiate] clear profit-sharing agreements or defining model ownership.'' Consequently, firms avoid collaborative AI development to prevent strategic disadvantages and ensure that they retain full control over their proprietary resources.

\section{Discussion}\label{sec:discuss}

In this work, we quantified and characterized the collaborative differences between OSS and OSM development paradigms, while investigating the underlying socio-technical factors through expert semi-structured interviews. Based on these insights, this section discusses the practical and theoretical implications of our work, as well as its limitations. Specifically, we discuss pathways to enhance collaboration in OSM and how our findings contribute to existing CSCW literature.

\subsection{Pathways to Enhance Collaboration in OSM}

To enhance collaborative practices in the OSM ecosystems, we explore pathways for improvement by both addressing identified barriers and leveraging emergent novel collaboration patterns, which may help practitioners navigate the unique constraints and opportunities within the OSM ecosystem.

\subsubsection{Operationalizing Transparency through Recipes}

Although full reproducibility in OSM is often hindered by resource constraints, transparency can be operationalized by standardizing training recipes~\cite{storks2023nlp}. Training recipes document data curation logic, hyperparameter configurations, and environment specifications, which can function as a standardized package to mitigate the resource gap between model producers and the broader community, thus supporting collective work. As P4 emphasized, ``We need a more sophisticated set of engineering practices to achieve one-click reproducibility \dots [such as] saving and sharing the whole recipe of the model.'' By decoupling the development logic from the underlying hardware requirements, it can provide a verifiable roadmap that allows the community to audit and build upon the work.

\subsubsection{Emulating Modularity to Lower Entry Barriers}

Emerging techniques such as Parameter-Efficient Fine-Tuning (PEFT)~\cite{hu2022lora} and model merging~\cite{yang2024model} serve as ``attempts to [emulate] modularity, paving the way for more decoupled development [workflows]''(P3). Such pseudo-modularity decouples the production of model variants from the end-to-end training process, enabling independent contributors to develop specialized functional adapters and synthesize them with base models to produce task-specific variants. By allowing for asynchronous refinement, these approaches lower entry barriers and reduce the heavy resource demands of pre-training.

\subsubsection{Constructing Model-Adapted Infrastructure}
Effective collaboration requires that infrastructure be specifically adapted to the socio-technical materiality of the collaborative artifact~\cite{star1994steps}. Since a critical challenge in current OSM collaborative development is the infrastructural misalignment, a model-adapted infrastructure that can manage large-scale binary weights and provide standardized, plug-and-use service may significantly promote collaboration in OSM~\cite{kulkarni2023llms}. As P7 noted, ``[W]e are using [fine-grained] algorithms like FastingCDC to attempt binary versioning for models \dots [to enable] differential storage and traceability.'' Furthermore, when integrated with ``standardized interfaces and [online] plug-and-use services, such as [standardized] inference engines like vLLM~\cite{kwon2023efficient} and automated ranking platforms''(P5), infrastructure facilitates quick feedback with less barriers, allowing independent contributors to test their innovations more effectively.

\subsubsection{Emerging Novel Collaborative Patterns}

We identified several emerging collaborative patterns within the model context. By leveraging these patterns, diverse forms of participation are enabled across the model lifecycle, thereby fostering better collaborative practices in OSM.

\textbf{\textit{Peripheral Knowledge Production.}} Contributors who are excluded from core model development due to significant resource and technical barriers still drive innovation~\cite{eder2019innovation}. These peripheral actors extend collaborative practice by ``creating knowledge commons that is independent from model structure [and weights], including the collective development of prompt libraries, evaluation benchmarks, and documentation''(P2). Without modifying the model itself, this practice empowers peripheral actors to enhance the community's collective intellectual property and promotes model utilization and governance.

\textbf{\textit{Collective Boundary Discovery.}} This pattern involves the collaborative effort to explore the behavioral and functional limits of an AI model by characterizing its unknown responses through distributed trials. Although the engagement of individuals is moving toward downstream usage, through collective experimentation~\cite{bjogvinsson2012design}, ``such as red-teaming and prompt test \dots contributors probe model [behavior] to identify bad cases and hallucination boundaries''(P10). These decentralized signals can guide upstream developers to refine model alignment in turn even when the core training process remains inaccessible.

\textbf{\textit{Resource-Based Collaboration.}} Unlike traditional software that enables collaboration through functional decomposition, the non-decomposable nature of AI models may enable collaboration by resource decoupling. As noted by P5, ``collaboration is organized through the exchange of complementary assets \dots large companies provide computational resource, while the partner companies [and community] contribute specialized expertise and domain-specific data.'' This exchange helps groups to get the external resources needed to overcome their limits~\cite{pfeffer1979external}, thereby building mutual dependencies and facilitating collaboration.

\subsection{Theoretical Contribution}

This research advances theoretical understanding of open source collaboration by examining how AI model development diverge from OSS patterns. We extend the literature by testing prior theory and contributing empirical findings from the context of OSM development.

\subsubsection{Testing Collaborative Theory}

Prior theory on OSS emphasizes commons-based peer production and collective intelligence~\cite{osborne2025characterising, benkler2006commons, raymond1999cathedral, o1999lessons}, where collaborative participation, inclusive processes and user innovation facilitate continuous improvement of shared artifacts through modular contributions. Our investigation empirically tests these assumptions in the OSM context and identifies significant distinctions that, besides substantially lower collaboration intensity and openness, collaboration in OSM development diverges from iteratively improving core components as \textit{``bazaar''} model~\cite{raymond1999cathedral} describes to resource complementarity and downstream adaptive usage. These findings reveal how a \textit{``bazaar''} model does not adequately capture OSM collaborative practices, where the nature of AI development (e.g., resource-intensive requirement and non-decomposability) reshapes collaborative dynamics and constrains participation opportunities.

\subsubsection{Extending Collaborative Development Frameworks}

Through quantitative findings and qualitative insights, we identified novel collaborative logic that extend CSCW frameworks. First, we shed light on a reconfiguration of core-periphery dynamics~\cite{borgatti2000models}. In OSS, the visibility of source code provides a degree of transparency and the potential for intervention, even if peripheral actors are not included in core development. In contrast, the structural inaccessibility of model training in OSM deepens the exclusion of peripheral actors. However, this exclusion does not lead to a decrease in peripheral labor. While peripheral work in OSS such as reporting issues often serves a corrective or auxiliary role, the OSM periphery becomes an important site for generative innovation~\cite{eder2019innovation}. Without access to core model development, value creation in OSM periphery shifts toward generating contextual knowledge, including fine-tuning protocols, alignment techniques, and usage experience, which constitutes a knowledge commons and provides substantial value for the broader developer ecosystem. Peripheral knowledge production therefore becomes a critical domain for innovation and contextualized application within OSM ecosystems.

Second, we noticed a potential transformation in articulation work~\cite{schmidt1992taking}. Articulation work in OSS focuses on coordinating distributed developers to achieve asynchronous integration and manage task allocation~\cite{boden2014articulation}. However, the non-decomposability of OSM redirects articulation work toward managing centralized resources as resource-based collaboration emerges. Coordinating access to computational resources, large-scale datasets, and human preference alignment processes constitutes the articulation work in OSM.

\subsubsection{Experimental and Infrastructural Participation}

Experimental practices~\cite{bjogvinsson2012design} emerge as a critical form of participation in OSM development. Although external developers rarely access core model development, they engage in collective experimental participation to map the capability boundaries and biases of black-box artifacts, redirecting distributed contribution pattern~\cite{chikersal2017deep} from constructing components to characterizing and delimiting behavior. These experimental activities generate feedback that indirectly informs model improvements, constituting a form of mediated and less visible collaboration. On the other hand, infrastructural participation~\cite{star1994steps} emerges as institutions provide computational resources without direct knowledge contributions, yet function as critical enablers for model development and collective experimentation. These findings introduce a new analytical perspective for understanding collaboration in OSM development beyond code-centric contribution models.

\subsection{Limitations and Future Work}

Despite the mixed-methods approach for comprehensiveness, we acknowledge several limitations in this work. First, in feature selection process, we limited our analysis to commits, issues/discussions, and stars/likes as indicators of developer communication and collaboration. While other activity data such as PRs also reflect developer interactions on GitHub~\cite{bertoncello2020pull}, we excluded them because the equivalent features are hardly used
by developers on the HF Hub, preventing cross-platform comparison.

Second, our study focused on two dominant open-source hosting platforms: GitHub (OSS) and HF Hub (OSM). Although other platforms exist in both domains, such as GitLab and Gitee for OSS, and TensorFlow Hub and PyTorch Hub for OSM, where collaborative practices may differ under different platform cultures, GitHub and HF Hub remain the most prominent in respective fields. Repositories from alternative platforms often maintain mirrors on or even migrate to these dominant platforms~\cite{cosentino2016findings, roveda2017does}. Moreover, our quantitative findings are further triangulated by qualitative insights from domain expert interviews, strengthening the validity of our results about development paradigm divergences between OSS and OSM.
However, platform-based interactions represent only one dimension of developer collaboration. Developers also engage through email, forums, social media, and face-to-face interactions~\cite{bertram2010communication, tamine2016social, sharma2022analyzing}, which are not captured by platform data mining. These alternative channels, which may exhibit different collaboration dynamics, call for further investigation to comprehensively understand the collaborative landscape in the context of OSM development.

Furthermore, despite adopting a sequential recruitment strategy based on thematic saturation principles, the limited sample size (N=10) may not capture the full spectrum of perspectives within the field. Additionally, the rapid evolution of AI community means that new viewpoints may have emerged subsequent to our investigation, potentially rendering some findings temporally constrained. Thus, further research is expected to validate, complement, and expand these findings across a broader AI development context.

On the other hand, given the fast-changing nature of AI community, longitudinal studies are expected to track how the identified collaborative patterns continue to evolve. Future research should also consider how infrastructural participation redistributes power and responsibility between institutions and communities, as well as exploring the design of model-adapted version control and collaborative infrastructure to enhance collaboration in OSM development.

\section{Conclusion}\label{sec:conclude}

This study presents a comprehensive investigation of the collaboration instensity, openness, and user innovation of OSS and OSM development paradigms. Employing a large-scale quantitative analysis of GitHub and HF Hub repositories, we found that compared to OSS development paradigm, the OSM development paradigm exhibits significantly lower collaboration intensity; lower collaboration openness regarding direct contribution while persisting relatively open knowledge exchange; and a divergence toward adaptive utilization user-innovation rather than collaborative improvement. Subsequent semi-structured interviews with domain experts revealed underlying socio-technical factors behind these distinctions, including architectural barriers, resource constraints, infrastructural misalignment and corporate strategic factors. 
By elucidating the divergence in distributed collaboration in the AI era, this study provides a foundation for developing targeted interventions.
We suggest future work to more comprehensively understand the collaborative practice across broader AI development context, and to explore infrastructure designs that can mitigate identified barriers for better collaboration in OSM development.


\begin{acks}
This work is supported by the Fundamental and Interdisciplinary Disciplines Breakthrough Plan of the Ministry of Education of China (No. JYB2025XDXM101) and the National Natural Science Foundation of China (No. 62332001). We thank the interview participants for their valuable time and insights, as well as the anonymous reviewers for their constructive feedback that helped improve this manuscript.
\end{acks}

\bibliographystyle{ACM-Reference-Format}
\bibliography{reference}

\appendix

\end{document}